\documentclass[twocolumn, amssymb, amsmath, aps, superscriptaddress, showpacs, footinbib,  prb]{revtex4}\usepackage{graphics}

\newcommand{\eff}{\mathrm{eff}}
\newcommand{\afm}{\mathrm{AFM}}

\begin{document}

\title{Magnetic properties of Ag$_2$VOP$_2$O$_7$: an unexpected spin dimer system}

\author{Alexander~A.~Tsirlin}
\email{altsirlin@gmail.com}
\affiliation{Max-Planck Institute CPfS, N\"othnitzer Str.~40, 01187 Dresden, Germany}
\affiliation{Department of Chemistry, Moscow State University, 119992 Moscow, Russia}
\author{Ramesh~Nath}
\author{Christoph~Geibel}
\author{Helge~Rosner}
\email{Helge.Rosner@cpfs.mpg.de}
\affiliation{Max-Planck Institute CPfS, N\"othnitzer Str.~40, 01187 Dresden, Germany}

\begin{abstract}
Magnetic properties of the silver vanadium phosphate Ag$_2$VOP$_2$O$_7$ are studied by means of magnetic susceptibility measurements and electronic structure calculations. In spite of the layered crystal structure suggesting 1D or 2D magnetic behavior, this compound can be understood as a spin dimer system. The fit of the magnetic susceptibility indicates an intradimer interaction of about 30 K in perfect agreement with the computational results. Our study emphasizes the possible pitfalls in interpreting experimental data on structural basis only and points out the importance of microscopic models for the understanding of the magnetic properties of vanadium phosphates. 
\end{abstract}

\maketitle

\section{Introduction}
Low-dimensional spin-1/2 systems have been extensively studied during the last decade due to strong quantum fluctuations that may lead to unusual ground states and low-temperature properties. Magnetic frustration in low-dimensional spin systems additionally suppresses long-range ordering and results in fascinating phenomena such as the formation of a resonating valence bond (RVB) spin-liquid ground state \cite{cav4o9,cav4o9-Pickett} or the magnetoelectrical effect in spin spirals.\cite{licu2o2,licuvo4} Recent studies \cite{cav4o9-Pickett,licu2o2-comment, licuvo4-properties,nacu2o2-Drechsler,licu2o2-Masuda} indicated the importance of combining experimental (measurement of thermodynamic properties, neutron and magnetic resonance studies) and computational (microscopic modeling via electronic structure calculations) approaches to gain an understanding of the unusual physical properties of low-dimensional spin-1/2 systems.

Vanadium phosphates present a promising and mainly unexplored field for the search of novel low-dimensional (and, possibly, frustrated) spin systems. Structural reports on vanadium phosphates are often accompanied by magnetic susceptibility data \cite{boudin} but few of the systems gained thorough physical investigation, the notable examples being (VO)$_2$P$_2$O$_7$ (Refs. \onlinecite{NIS-pyrophosphate,NMR-pyrophosphate,high-pressure}) and VO(HPO$_4)\cdot 0.5$H$_2$O.\cite{semihydrate,NIS-semihydrate} The careful study of these compounds revealed complex superexchange pathways and showed the failure of the straightforward interpretation of experimental results on the structural basis only.\cite{NIS-pyrophosphate,NIS-semihydrate}

Recently, we studied magnetic properties of a novel vanadium phosphate Pb$_2$VO(PO$_4$)$_2$ and found an interesting realization of the frustrated square lattice system with both ferro- and antiferromagnetic interactions resulting in columnar antiferromagnetic ordering.\cite{kaul,roms,skoulatos} Thus, vanadium phosphates can reveal very unusual spin systems, and a detailed study of the respective compounds is of high interest. Below, we present the investigation of the magnetic properties of a silver vanadium phosphate Ag$_2$VOP$_2$O$_7$.\cite{ag2vp2o8} We use both experimental and computational techniques in order to achieve a reliable description of the exchange couplings in the system under investigation. The results are discussed with respect to the structural correlations for magnetic interactions in vanadium phosphates.

The outline of the paper is as follows. In Sec. \ref{structure}, we briefly describe the crystal structure and analyze the possible pathways of superexchange interactions. Methodological aspects are given in Sec. \ref{method}. Sec. \ref{experimental} deals with magnetic susceptibility data, while Sec. \ref{band} presents the results of band structure calculations and estimates for the exchange integrals. Experimental and computational results are compared and discussed in Sec. \ref{discussion} followed by our conclusions. 

\section{Crystal structure}
\label{structure}
\begin{figure}
\includegraphics{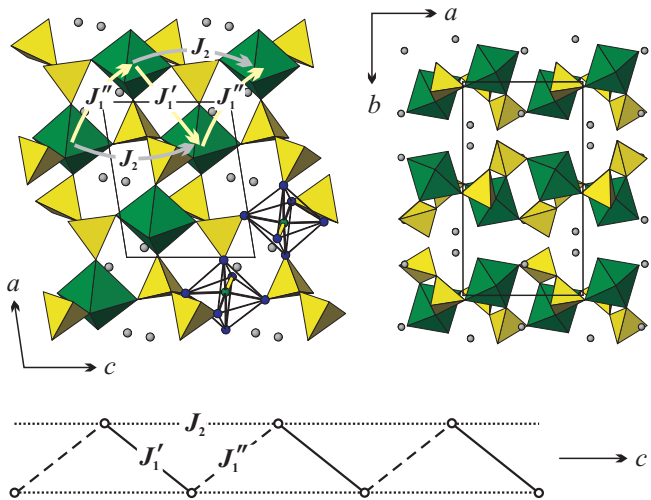}
\caption{\label{fig-structure}(Color online) Crystal structure of Ag$_2$VOP$_2$O$_7$ (upper panels) and a schematic image of an alternating zigzag chain (bottom panel). The upper left panel shows a single [VOP$_2$O$_7$] layer, while the upper right panel presents the stacking of the layers. In the upper left panel, faces of the two VO$_6$ octahedra are removed in order to illustrate the orientation of short V--O bonds (shown as thicker lines). Solid, dashed, and dotted lines in the bottom panel denote $J_1',\ J_1''$, and $J_2$, respectively.}
\end{figure}

The crystal structure of Ag$_2$VOP$_2$O$_7$ is formed by [VOP$_2$O$_7$] layers separated by silver cations. The structure has monoclinic symmetry ($P2_1/c$, $a=7.739$ \r A, $b=13.611$~\r A, $c=6.294$ \r A, $\beta=99.0^0$, $Z=4$). One unit cell contains two layers and, in particular, four vanadium atoms -- two from each of the layers (Fig. \ref{fig-structure}). Every layer includes VO$_6$ octahedra joined by PO$_4$ tetrahedra. Vanadium has the oxidation state of +4 corresponding to the electronic configuration $d^1$, i.e., $S=1/2$.\cite{ag2vp2o8} Magnetic measurements down to 77 K indicate paramagnetic behavior of Ag$_2$VOP$_2$O$_7$,\cite{ag2vp2o8} but, as we will show in the following, considerable deviation from the Curie law starts immediately below this temperature.

To understand the possible magnetic interactions in Ag$_2$VOP$_2$O$_7$, one has to consider the local environment of vanadium atoms as well as the connections between vanadium polyhedra. The VO$_6$ octahedra are strongly distorted due to the formation of a short V--O bond typical for V$^{+4}$ (see Ref. \onlinecite{boudin}). The distortion of the octahedron results in a non-degenerate orbital ground state with the half-filled $d_{xy}$ orbital lying in the equatorial plane of the octahedron,\cite{sr2v3o9} i.e., almost parallel to the [VOP$_2$O$_7$] layers. 

The orientation of the half-filled orbital implies that interlayer interactions should be negligibly small while the magnitude of in-layer interactions depends on the connections between the vanadium octahedra. According to Fig. \ref{fig-structure}, V--O--V superexchange is impossible in Ag$_2$VOP$_2$O$_7$; therefore, all the superexchange pathways must include PO$_4$ tetrahedra. Assuming that the interactions via two consecutive tetrahedra (i.e., the pyrophosphate P$_2$O$_7$ group) are weak, we derive three possible couplings: $J_1',\ J_1''$, and $J_2$ indicating a 1D spin system. One can consider such a system as an alternating zigzag chain (see bottom panel of Fig. \ref{fig-structure}), since two nearest-neighbor interactions $J_1',\ J_1''$ are not equivalent by symmetry, while $J_2$ is the next-nearest neighbor interaction. Interchain couplings are weak as they include P$_2$O$_7$ groups or single tetrahedra that do not match the half-filled orbital of vanadium. Below (Sec. \ref{band}), we justify the weakness of the interchain couplings by analyzing the local density approximation (LDA) band structure with a tight-binding model that accounts for \emph{all} the interactions between nearest and next-nearest neighbors.

\section{Methods}
\label{method}
Polycrystalline samples of Ag$_2$VOP$_2$O$_7$ were prepared according to the procedure reported in Ref. \onlinecite{ag2vp2o8}. Stoichiometric amounts of AgNO$_3$, (NH$_4)_2$HPO$_4$, and VO$_2$ were carefully mixed, placed in corundum crucibles and annealed in air. The regime of the annealing was as follows: 12 hours at 200 $^0$C, 1 day at 400 $^0$C, and 1 day at 500 $^0$C with regrindings after each step. The final product was a single phase as checked by X-ray diffraction (XRD) (Huber G670f camera, CuK$_{\alpha1}$ radiation, ImagePlate detector). However, the XRD pattern revealed a rather high background signal that probably indicated low crystallinity of the prepared samples or the presence of amorphous impurities. Unfortunately, we failed to improve the quality of the samples, since further annealings (in air or in vacuum) did not lead to any notable changes in the XRD patterns or resulted in the appearance of non-identified impurity peaks. We also tried to prepare an isostructural compound Na$_2$VOP$_2$O$_7$ (Ref. \onlinecite{na2vp2o8}) but the samples always contained a mixture of two polymorphic modifications [one is isostructural to Ag$_2$VOP$_2$O$_7$ while the other one has fresnoite-type structure (see Ref.~\onlinecite{na2vop2o7})].

Magnetic susceptibility was measured using Quantum Design SQUID between 2 and 300 K in the fields $\mu_0H$ of 0.5, 1, and 5 T. 

Scalar-relativistic density-functional (DFT) band structure calculations were performed using a full-potential local-orbital scheme\cite{fplo} (FPLO5.00-19) and the exchange-correlation potential of Perdew and Wang.\cite{perdew} Ag$(4s,4p,4d,5s,5p)$, V$(3s,3p,3d,4s,4p)$, P$(2s,2p,3s,3p,3d)$, and O$(2s,2p,3d)$ orbitals were employed as the basis set, while all lower-lying orbitals were treated as core states. The $k$ mesh included 864 points within the first Brillouin zone (296 in its irreducible part). The convergence with respect to the number of $k$ points was carefully checked. 

First, a LDA calculation was performed employing the crystallographic unit cell and the full symmetry of the crystal structure ($P2_1/c$). Orbital states relevant for magnetic interactions were chosen on the basis of the LDA band structure and the respective bands were analyzed using a tight-binding model. Such analysis provided an estimate for the magnitudes of all nearest-neighbor and next-nearest-neighbor exchange couplings in Ag$_2$VOP$_2$O$_7$. We found that the values of interest were $J_1',\ J_1'',\ J_2$ (see Fig.~\ref{fig-structure}), in agreement with the empirical structural considerations listed in Sec. \ref{structure}. 

Next we turned to the LSDA+$U$ technique in order to calculate total energies for several patterns of spin ordering and to give an independent estimate of the leading exchange interactions in Ag$_2$VOP$_2$O$_7$. Such an estimate requires doubling of the lattice translation in the $c$ direction (see Section \ref{band}). The doubling of the crystallographic unit cell results in $Z=8$ and 104 atoms making the calculations very time-consuming and less accurate. Therefore, we slightly simplified the crystal structure in our LSDA+$U$ calculations. The simplification was justified by the similarity of LDA band structures and tight-binding models for the crystallographic unit cell and the supercell (i.e., the modified cell).

The neighboring layers in Ag$_2$VOP$_2$O$_7$ are not translationally equivalent, but they \emph{are} equivalent by symmetry, hence they reveal identical in-layer interactions. Interlayer interactions are very weak (see Sec. \ref{band}), therefore the unit cell with one layer is actually sufficient for the present computational purpose. We use a $P1$ supercell with $a_s=a=7.739$ \r A, $b_s=8.5069$ \r A ($\sim 0.6b$), $c_s=2c=12.588$ \r A, and $\beta_s=\beta=99^0$. The supercell includes one [VOP$_2$O$_7$] layer that \emph{exactly} matches the layer in the real structure. However, the local environment of silver atoms is changed, and the $b_s$ parameter is slightly elongated as compared to $b/2$ in order to avoid unreasonably short Ag--O distances. The modification of the crystal structure is a strong but \emph{well justified} approximation, since silver atoms do not take part in superexchange pathways, while the geometry of the V--P--O framework (crucial for magnetic interactions) is unchanged. These conclusions are supported by similar tight-binding fits of the relevant bands in LDA band structures for the crystallographic unit cell and the supercell. The variation of the respective hopping parameters does not exceed 5\%, hence the error for $J$'s is below 10\% (i.e., within the typical accuracy of LSDA+$U$).

Another problematic point of LSDA+$U$ calculations deals with the choice of the Coulomb repulsion parameter $U_d$ [we use the notation $U_d$ in order to distinguish this parameter of the computational method from the model parameter $U_{\eff}$ for the Coulomb repulsion in the effective one-band model (see Sec. \ref{band})]. Below, we use several physically reasonable values of $U_d$ (namely, 4, 5, and 6 eV) in order to check the influence of $U_d$ on the resulting exchange integrals. The exchange parameter of LSDA+$U$ is fixed at $J=1$ eV, as usually it has no significant influence on the results.

\section{Experimental results}
\label{experimental}
\begin{figure}
\includegraphics{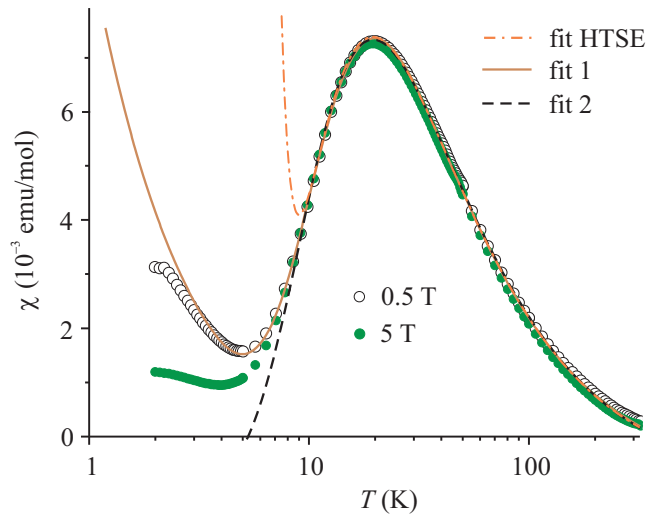}
\caption{\label{fig-suscept}(Color online) Magnetic susceptibility of Ag$_2$VOP$_2$O$_7$ measured at fields $\mu_0H$ of 0.5 T (empty circles) and 5 T (filled circles). Solid and dashed lines show the fits with the model of $S=1/2$ alternating chain according to the first and second lines of Table \ref{table-fit}, respectively. The dash-dotted line is the fit with the HTSE (Ref. \onlinecite{HTSE}) above 10 K.}
\end{figure}

Magnetic susceptibility curves for Ag$_2$VOP$_2$O$_7$ are shown in Fig. \ref{fig-suscept}. The curves reveal a maximum at $T_{max}\approx 20$ K and a fast decrease from $T_{max}$ down to 5 K indicating spin-gap-like behavior. Below 5 K the susceptibility increases (at low fields) or remains nearly temperature-independent (at 5 T). The samples under investigation are powders, therefore the low-temperature part of the data is dominated by the contribution of paramagnetic impurities and defects saturated at high field. The susceptibility curves above 2.2 K do not show any anomalies that could denote long-range magnetic ordering.\cite{foot1} A weak bend is present on the low-field curve below 2.2 K, but the present data (collected above 2 K) are not sufficient to reveal the origin of this bend. 

Thus, the susceptibility data for Ag$_2$VOP$_2$O$_7$ reveal low-dimensional spin-gap-like behavior consistent with the scheme of magnetic interactions proposed in Sec.~\ref{structure}. The alternation of nearest-neighbor interactions $J_1'$ and $J_1''$ causes the formation of the spin gap. The high-temperature part of the susceptibility (above 60 K) shows the Curie-Weiss type behavior, although the $1/\chi$ \textit{vs.} $T$ curves are nonlinear due to considerable temperature-independent contribution. We fit the data above 100 K as $\chi=\chi_0+C/(T-\theta)$ and find $\chi_0=-7.9(1)\cdot 10^{-4}$ emu/mol, $C=0.350(1)$ emu$\cdot$K/mol, and $\theta=11.0(4)$~K. The Curie constant $C$ corresponds to the effective moment of $1.62\ \mu_B$ that is slightly lower than the expected spin-only value of $1.73\ \mu_B$. The discrepancy may be caused by the presence of nonmagnetic impurities in the samples under investigation, and this explanation is additionally supported by high diamagnetic contribution $\chi_0$ as well as by the slightly decreased $g$ value (see below). We suggest that the high diamagnetic contribution is caused by the preparation procedure, since the samples are heated in air, i.e., in oxidative conditions favoring the formation of diamagnetic V$^{+5}$. Unfortunately, annealing in vacuum did not result in single-phase samples. 

Now, we fit the experimental data with model expressions. Basically, one should use the model of the alternating zigzag chain, as suggested by structural considerations in Sec. \ref{structure}. However, only high-temperature series expansion (HTSE) results are available for the susceptibility of this model.\cite{HTSE} We find a perfect fit of the experimental data above 10 K (see Fig. \ref{fig-suscept}) with $g=1.92(1)$, $J_1'=31.3(1)$ K, $J_1''=0.8(3)$ K, $J_2=5.0(6)$ K, and $\chi_0=-8.7(1)\cdot 10^{-4}$ emu/mol (temperature-independent contribution). Note that we have assumed $J_1''<J_1'$: at the stage of fitting, this choice is, of course, arbitrary, and one may invert the assumption without any change in the results. One should not be surprised by the use of the HTSE for temperatures well below $J_1'$, as the expansion under consideration is valid down to $T\approx 0.25J$ (see Ref. \onlinecite{HTSE}). Moreover, the data below the susceptibility maximum (i.e., below 20 K) are crucial for the stability of the fitting and the obtaining of reasonable exchange couplings. 

Unfortunately, the HTSE is inapplicable in the low-temperature region. To fit the decrease of the susceptibility between 20 and 5 K, we have to turn to a simplified model. If one neglects $J_2$, Ag$_2$VOP$_2$O$_7$ becomes an alternating chain system. The experimental data above 5~K are fitted with the expression:
\begin{equation}
  \chi=\chi_0+C_i/T+\chi_{alt.chain},
\label{eq-fit}\end{equation}
where $\chi_0$ is the temperature-independent contribution, $C_i/T$ accounts for the Curie-like paramagnetism of impurities and defects, while $\chi_{alt.chain}$ is the susceptibility of an $S=1/2$ alternating chain as given in Ref. \onlinecite{johnston}. 

The term $\chi_{alt.chain}$ includes three variable parameters: $g,\ J_1$, and the alternation coefficient $\alpha$. Again, we assume $J_1''<J_1'$, then $J_1=J_1'$ and $\alpha=J_1''/J_1'$.  We find a perfect fit of the experimental data above 5 K (fit 1 in Fig. \ref{fig-suscept}) with the parameters listed in the first row of Table \ref{table-fit}. Below 5 K, the experimental data deviate from the model due to the deviation of the paramagnetic impurity contribution from the Curie law. Note that one may skip the $C_i/T$ term and find a reasonable fit above 8 K (fit 2 at Fig. \ref{fig-suscept}; the parameters are listed in the second row of Table \ref{table-fit}), but the data between 5 and 8 K are poorly described. However, the second fit presents a somewhat better $g$ value that is close to the expected range of 1.93--1.96 (see Refs. \onlinecite{livpo5} and \onlinecite{sr2v3o9-2}). On the other hand, $g$ may be decreased due to the presence of nonmagnetic impurities in the sample under investigation (as shown by the Curie-Weiss fit above) or due to the neglect of $J_2$. 

\begin{table}
\caption{\label{table-fit}The results of the fitting of magnetic susceptibility data with an alternating chain model [eq. (\ref{eq-fit})]. $\chi_0$ is a temperature-independent contribution, $C_i$ denotes the Curie-like $C_i/T$ contribution, $g$ is the $g$-factor, $J_1$ labels the leading in-chain interaction, and $\alpha$ is the alternation parameter. Fit 1 and fit 2 are marked according to Fig. \ref{fig-suscept}.}
\begin{ruledtabular}
\begin{tabular}{@{}cccccc}
& $\chi_0$ (emu/mol) & $C_i$ (emu$\cdot$K/mol) & $g$ & $J_1$ (K) & $\alpha$ \\
Fit 1 & $-8.3(1)\cdot 10^{-4}$ & $0.0100(2)$ & $1.86(1)$ & $33.1(1)$ & $0.12(2)$ \\
Fit 2 & $-9.0(1)\cdot 10^{-4}$ & --- & $1.92(1)$ & $31.6(1)$ & $0.31(1)$ \\
\end{tabular}
\end{ruledtabular}
\end{table}

Thus, the present experimental data leave some uncertainty in the $\alpha$ value. In our opinion, one can not neglect the Curie-like paramagnetic contributions while dealing with the low-temperature susceptibility of polycrystalline samples. Therefore, we suggest that the first fit with the extremely low $\alpha$ is more realistic as confirmed by the HTSE fit [and band structure calculations (see the next section)]. Note that both fits reveal a considerable difference between $J_1'$ and $J_1''$ (at least by a factor of 3). However, the scenario of $J_1''\ll J_1'$ is rather unexpected, since superexchange pathways of the two interactions are very similar. 

\section{Band structure}
\label{band}
\begin{figure}
\includegraphics{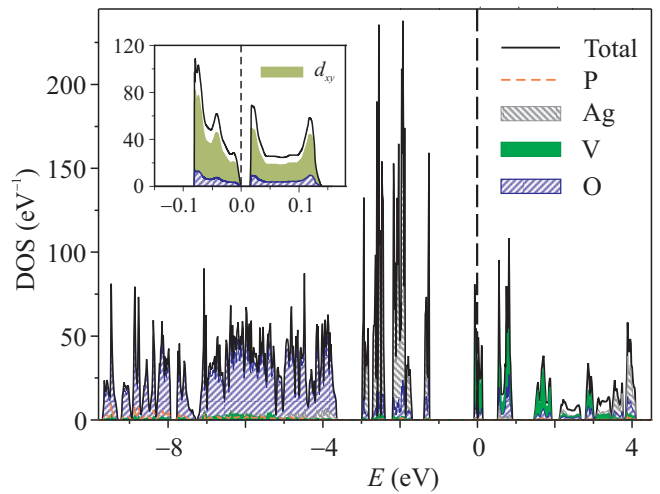}
\caption{\label{fig-dos}(Color online) Total and atomic resolved LDA density of states for Ag$_2$VOP$_2$O$_7$. The Fermi level is at zero energy. The inset shows the blowup of the image near the Fermi level. The solid filling in the primary figure marks the contribution of \emph{all} vanadium orbitals, while the solid filling in the inset corresponds to V $3d_{xy}$ orbitals \emph{only}.}
\end{figure}
The plot of LDA density of states is shown in Fig.~\ref{fig-dos}. At large, the band structure of Ag$_2$VOP$_2$O$_7$ is similar to that of other vanadium compounds (see, for instance, Refs. \onlinecite{sr2v3o9} and \onlinecite{korotin}): the states below $-3.5$ eV are dominated by oxygen orbitals, while the states near the Fermi level have mainly vanadium $3d$ character with minor oxygen contribution. However, a set of narrow bands between $-1$ and $-3$ eV is a distinctive feature of Ag$_2$VOP$_2$O$_7$. These bands correspond to $4d$ orbitals of silver and are completely filled as one would readily expect for Ag$^{+1}$ ($4d^{10}$). The LDA energy spectrum reveals a narrow gap of about 0.015 eV that is definitely too small to account for the green color of Ag$_2$VOP$_2$O$_7$. The strong underestimate (or even the lack) of the energy gap is a typical failure of LDA in transition metal compounds due to an improper description of the correlation effects in the $3d$ shell (especially for half-filled orbitals, such as $3d_{xy}$ in V$^{+4}$). Realistic insulating spectra with $E_g=2.0-2.3$ eV are readily obtained by means of LSDA+$U$ (see below).

Below, we analyze the band structure in order to find the values of exchange couplings and to construct a reliable model of magnetic interactions in Ag$_2$VOP$_2$O$_7$. First, we focus on the LDA bands near the Fermi level and extract transfer integrals for vanadium orbitals using a tight-binding model. The transfer integrals ($t$) are introduced to the extended Hubbard model, and the correlation effects are taken into account explicitly via an effective on-site repulsion potential $U_{\eff}$. In the strongly correlated limit $t\ll U_{\eff}$, the half-filled Hubbard model can be reduced to a Heisenberg model for the low-lying excitations. Thus, we estimate magnitudes of all exchange interactions between nearest and next-nearest neighbors. Next, we give an independent estimate of several exchange coupling constants by introducing the correlation effects in a mean-field approximation to the band structure (LSDA+$U$) and calculating total energies for different patterns of spin ordering.

\begin{figure}
\includegraphics{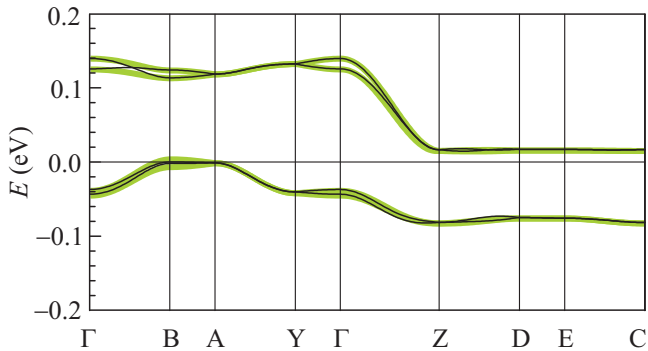}
\caption{\label{fig-bands}(Color online) LDA band structure of Ag$_2$VOP$_2$O$_7$ near the Fermi level. Thick (green) lines show the fit of the tight-binding model. The Fermi level is at zero energy. The notation of $k$ points is as follows: $\Gamma(0,0,0)$, B$(0.5,0,0)$, A$(0.5,0.5,0)$, Y$(0,0.5,0)$, Z$(0,0,0.5)$, D$(0.5,0,0.5)$, E$(0.5,0.5,0.5)$, C$(0,0.5,0.5)$ (the coordinates are given along $k_x,\ k_y$, and $k_z$ axes in units of the respective reciprocal lattice parameters).} 
\end{figure}

The four bands near the Fermi level have V $d_{xy}$ character (see the inset Fig. \ref{fig-dos}), in agreement with the qualitative discussion in Sec. \ref{structure}. We fit a four-band tight-binding model to these bands (Fig. \ref{fig-bands}) and use the resulting hopping parameters to estimate antiferromagnetic contributions to the exchange integrals as $J_i^{\afm}=4t_i^2/U_{\eff}$. $U_{\eff}$ is the effective on-site Coulomb repulsion. According to Ref. \onlinecite{sr2v3o9}, $U_{\eff}$ lies in the range of $4-5$ eV for vanadium oxides, and we use $U_{\eff}=4.5$~eV in the following discussion as a representative value. Note that the change of $U_{\eff}$ results in a simple scaling of the exchange integrals, therefore the ratios of $J$'s remain constant. 

\begin{table}
\caption{\label{table-tb}Leading hopping parameters of the tight-binding model and the resulting values of $J_i^{\afm}$}
\begin{ruledtabular}
\begin{tabular}{@{}cccc}
& $t_1'$ & $t_1''$ & $t_2$ \\
$t$ (eV) & 0.060 & 0.013 & $-0.021$ \\\hline
& $J_1'$ & $J_1''$ & $J_2$ \\
$J^{\afm}$ (K) & 37 & 1.8 & 4.6 \\
\end{tabular}
\end{ruledtabular}
\end{table}
The tight-binding model included seven in-layer hoppings (i.e., all nearest-neighbor and next-nearest-neighbor hoppings) and an interlayer hopping $t_{\perp}$. The resulting $t_{\perp}\approx 0.001$ eV is very small, as one can readily see from Fig. \ref{fig-bands}: the bands are close to double degeneracy since two vanadium atoms of one layer weakly interact with the respective atoms of the neighboring layer. The leading in-layer hoppings $t_1',\ t_1''$, and $t_2$ (see Fig. \ref{fig-structure}) are listed in Table \ref{table-tb}, while other in-layer hoppings are small (below 0.008 eV) and can be neglected. Thus, the analysis of the LDA band structure confirms the empirical structural considerations listed in Sec. \ref{structure}.

The leading transfer integral $t_1'$ corresponds to $J_1'^{\afm}\approx 37$ K, while $J_1''^{\afm}$ and $J_2^{\afm}$ are weaker by an order of magnitude. This result is in a remarkable agreement with the experimental data matching both the value of the strongest coupling and the striking difference between nearest-neighbor interactions $J_1'$ and $J_1''$. Note, however, that the tight-binding model does not allow us to distinguish between two structurally similar superexchange pathways, and one can also fit the bands with $t_1''\gg t_1'$. Now, we proceed to a different computational technique that helps us to resolve $t_1',\ t_1''$ and provides additional confirmation for the difference of the respective exchange integrals.

\begin{table}
\caption{\label{table-integrals}Exchange integrals calculated with LSDA+$U$}
\begin{ruledtabular}
\begin{tabular}{@{}cccc}
$U_d$ (eV) & $J_1'$ (K) & $J_1''$ (K) & $J_2$ (K) \\
4 & 51.4 & 1.0 & 8.1 \\
5 & 47.3 & 4.1 & 5.7 \\
6 & 41.4 & 4.6 & 4.4 \\
\end{tabular}
\end{ruledtabular}
\end{table}

\begin{figure}
\includegraphics{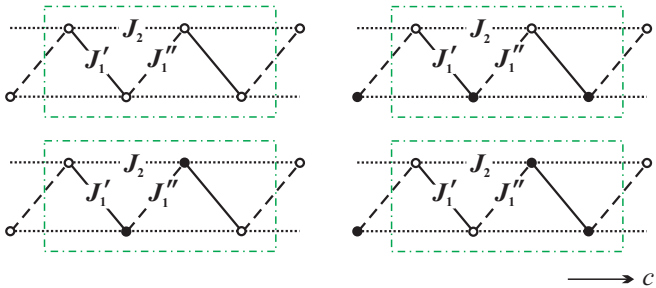}
\caption{\label{fig-patterns}(Color online) Four patterns of spin ordering employed for the calculation of $J$'s with LSDA+$U$. Dash-dotted rectangles show the repetition unit (i.e. the supercell). Circles denote vanadium atoms; full and empty circles mark different spin directions.} 
\end{figure}

The hopping parameters (indicating the width of the half-filled bands) are two orders of magnitude smaller than the Coulomb repulsion; therefore, we may use the LSDA+$U$ technique to account for correlation effects within band structure calculations in a mean-field approximation. Four patterns of spin ordering (see Fig. \ref{fig-patterns}) were used in order to estimate $J_1',\ J_1''$, and $J_2$. The resulting exchange integrals are listed in Table \ref{table-integrals}. The absolute values of $J$'s depend on $U_d$ considerably, but qualitatively the results are similar. We find that $J_1'$ is indeed the strongest coupling, it exceeds $J_1'',\ J_2$ at least by a factor of 6. $J_1'$ is the interaction corresponding to longer V--V separation (5.29 \r A \textit{vs.} 4.92 \r A for $J_1''$) as shown in Fig. \ref{fig-structure}. The small value of $J_2$ is consistent with the tight-binding results and provides a further justification for the fitting of experimental data with the alternating chain model (see Sec. \ref{experimental}). All the interactions in Ag$_2$VOP$_2$O$_7$ are antiferromagnetic.

As we have mentioned in Sec. \ref{method}, the choice of $U_d$ is a problematic point of the LSDA+$U$ approach. The $U_d$ parameter is usually fitted to experimentally observed properties (e.g., energy gaps, magnetic moments) or estimated by means of constrained LDA calculation (see, for example, Ref. \onlinecite{korotin}). Basically, the choice of $U_d$ remains somewhat ambiguous and depends on the particular experimentally observed quantities used for the fitting as well as structural features and the computational method.

The results for $U_d=6$ eV show the best agreement with the experimental and tight-binding estimates of $J$'s. Note also that the largest energy gap of 2.3 eV found for $U_d=6$ eV seems to be consistent with the green color of Ag$_2$VOP$_2$O$_7$. These findings are somewhat unexpected since we used $U_{\eff}=4.5$ eV, and even smaller $U_d=3.6$ eV had been applied in LSDA+$U$ calculations for vanadium oxides before.\cite{korotin} However, one should always keep in mind that $U_{\eff}$ and $U_d$ are different quantities having different physical meaning. $U_{\eff}$ is an effective repulsion in a mixed vanadium-oxygen band, while the repulsive potential $U_d$ is applied to V $3d$ orbitals only. Thus, $U_d$ should exceed $U_{\eff}$, since $U_{\eff}$ corresponds to "molecular"\ V--O orbitals having larger spatial extension as compared to that of atomic V $3d$ orbitals. As for the previous LSDA+$U$ calculations for vanadium oxides,\cite{korotin} the difference in computational methods and crystal structures may be relevant. The linearized muffin-tin orbital (LMTO) basis is used in Ref. \onlinecite{korotin}, while we apply a different basis of local orbitals. Moreover, Korotin \textit{et al.}\cite{korotin} studied layered compounds containing V$^{+4}$ in square-pyramidal coordination, while vanadium is surrounded by six oxygen atoms in Ag$_2$VOP$_2$O$_7$. Thus, the difference in the optimal $U_d$ value is plausible, although the origin of this difference is not completely clear.

\section{Discussion}
\label{discussion}
Experimental data and band structure calculations provide a consistent microscopic scenario for the magnetic interactions in Ag$_2$VOP$_2$O$_7$. One of the nearest-neighbor couplings ($J_1'$) is about 30 K and exceeds the other couplings at least by a factor of 6. Thus, Ag$_2$VOP$_2$O$_7$ may be considered as a system of weakly coupled dimers. Indeed, a simple model of isolated dimers [with additional terms $\chi_0+C_i/T$ similar to eq. (\ref{eq-fit})] with $g=1.83$ and $J=33$ K perfectly fits the experimental data above 5 K. This fit completely matches fit 1 in Fig. \ref{fig-suscept}. Interdimer interactions $J_1''$ and $J_2$ may be responsible for the slight variation of the $g$ value, and they certainly influence the possible onset of the long-range ordering. The HTSE fit and band structure calculations indicate that both $J_1''$ and $J_2$ are antiferromagnetic, therefore frustration should be expected. However, it will not change the main feature of Ag$_2$VOP$_2$O$_7$ -- a spin-gap-type magnetic behavior due to strong intradimer coupling.

The identification of weakly coupled spin dimers in Ag$_2$VOP$_2$O$_7$ is surprising, since two nearest-neighbor interactions -- $J_1'$ and $J_1''$ -- have very similar superexchange pathways. The differences in interatomic distances and angles for these V--O--P--O--V connections do not exceed 0.05 \r A and $10^{\circ}$, respectively. Nevertheless, the tiny structural changes strongly influence the values of the exchange integrals. 

As we have mentioned in the introduction, superexchange pathways in vanadium phosphates are sometimes non-straightforward. However, some clarity may be achieved by considering two assumptions. First, V--O--P--O--V superexchange pathways are often much more efficient than V--O--V ones. Second, the magnitude of interactions involving PO$_4$ tetrahedra can be estimated using structural parameters [via the so-called magnetostructural correlations, see Refs. \onlinecite{roca} and \onlinecite{petit}]. These correlations suggest that double phosphate bridges (i.e., two PO$_4$ tetrahedra linking two VO$_6$ octahedra) are usually more efficient compared to single bridges, while the interactions via double bridges depend on the relative orientation of vanadium and phosphorous polyhedra quantified by several geometrical parameters. The present study may provide a good test for such structural considerations concerning magnetic interactions in systems with complex superexchange pathways.

In case of Ag$_2$VOP$_2$O$_7$, double phosphate bridges are relevant for both $J_1'$ and $J_1''$. Most geometrical parameters are similar for these interactions, and the only difference deals with the in-plane displacement of the VO$_6$ octahedra (plane implies the equatorial plane of octahedra, i.e., the plane of the half-filled $d_{xy}$ orbital). Such displacement favors $J_1'$ rather than $J_1''$, although one would not expect a considerable difference between the two integrals. However, our experimental and computational results suggest the scenario of $J_1'/J_1''>6$ resulting in a system of weakly coupled dimers rather than in an alternating chain. Basically, the structural considerations suggest correct trends, but quantitative estimates require more sophisticated techniques.

In this paper, we tried to use an adequate computational approach to the study of vanadium phosphates. All the computational results available are based on very simplified techniques such as extended H\"uckel \cite{roca,pyrophosphate-2000,whangbo-2002} or DFT GGA calculations \cite{petit,semihydrate-2004,pyrophosphate-2002} for small clusters (typically, spin dimers -- two VO$_6$ octahedra and bridging PO$_4$ tetrahedra). Now, we show that vanadium phosphates are readily accessible to full-potential band structure calculations that are known as a very efficient tool for the study of magnetic properties of transition metal oxides.\cite{cav4o9-Pickett, korotin, sr2v3o9,licu2o2-comment,licuvo4-properties,nacu2o2-Drechsler,li2vosio4,li2cuzro4} Full-potential calculations provide reasonable accuracy and set the computational results on much safer grounds compared to cluster calculations, since the crude cluster approximation is avoided and correlation effects (which are very important for the properties of transition metal compounds) can be included. We are looking forward to the further application of full-potential calculations to vanadium phosphates and to the discovery of novel unusual spin systems in these compounds.

It is worth noting that Ag$_2$VOP$_2$O$_7$ may be an interesting realization of the alternating zigzag chain model. This model, also known as dimerized zigzag chain or dimerized frustrated chain, has been extensively studied theoretically with a particular interest to its excitation spectrum.\cite{chitra-1995,excitations-1996,excitations-1999,excitations-1998,excitations-2001,excitations-2004} However, the experimental access to the model is still scarce and basically limited by the spin Peierls phase of CuGeO$_3$ that corresponds to the weakly dimerized case. As we have stated above, Ag$_2$VOP$_2$O$_7$ is strongly dimerized, and the dominating spin-gap properties are observed. Nevertheless, the frustration of weak interdimer couplings may be manifested in high-field properties of the material or its magnetic excitations. 

Using the representative values of $J_1'\simeq 30$ K, $J_1''\simeq J_2\simeq 5$ K, we find $\alpha_f\simeq 0.17$, $\delta\simeq 0.7$ and estimate the position of Ag$_2$VOP$_2$O$_7$ on the phase diagram of the model [$\alpha_f=J_2/J_1'$ is the frustration ratio, while $\delta=(J'-J'')/(J'+J'')$ is the dimerization ratio]. According to Refs.~\onlinecite{magnetization} and \onlinecite{magnetization-2}, this position implies the presence of magnetization plateau, while Ref.~\onlinecite{excitations-1999} suggests the formation of bound singlet and triplet states above the elementary triplet excitation that corresponds to the spin gap. Thus, a detailed study of the high-field properties of Ag$_2$VOP$_2$O$_7$ and neutron scattering experiments probing the magnetic excitations of this compound could be interesting.

In conclusion, we studied magnetic properties of Ag$_2$VOP$_2$O$_7$ using magnetic susceptibility measurements and full-potential band structure calculations. We reach a perfect consistency between experimental and computational results and describe Ag$_2$VOP$_2$O$_7$ as a system of weakly coupled dimers on the alternating zigzag chain, although the strong dimerization is rather surprising from the structural point of view. The frustration of interdimer interactions may lead to a number of unusual properties that deserve a further experimental study. We present the first application of full-potential band structure calculations to vanadium phosphates. The calculations show reasonably accurate and highly reliable results, hence they should be an efficient tool in the study of related compounds.

\section*{Acknowledgements}
The authors acknowledge financial support of RFBR (Project No. 07-03-00890), GIF (Grant No. I-811-257.14/03), and the Emmy-Noether-Program of the DFG. ZIH Dresden is acknowledged for computational facilities. We are grateful to Evgeny Antipov for careful reading of the manuscript and valuable comments. A.Ts. is also grateful to MPI CPfS for hospitality and financial support during the stay.


\begin{thebibliography}{45}
\bibitem[{foo()}]{foot1}
\bibinfo{note}{Nevertheless, one should not make a decisive conclusion about
  long-range magnetic ordering in Ag$_2$VOP$_2$O$_7$ on the basis of
  susceptibility data only, since transition anomalies are usually weak in
  polycrystalline samples (see, for instance, Ref. \onlinecite{kaul}) and may
  be additionally concealed by the low-temperature paramagnetic upturn.}
\end{thebibliography}
\end{document}